# Reconfigurable SDM Switching Using Novel Silicon Photonic Integrated Circuit

Yunhong Ding,[1*] Valerija Kamchevska,[1] Kjeld Dalgaard,[1] Feihong Ye,[1] Rameez Asif,[1] Simon Gross,[2] Michael J. Withford,[2] Michael Galili,[1] Toshio Morioka,[1] and Leif Katsuo Oxenløwe[1]

[1] DTU Fotonik, Department of Photonics Engineering, Technical University of Denmark, Ørsteds Plads 343, DK-2800 Kgs. Lyngby, Denmark
[2] Centre for Ultrahigh bandwidth Devices for Optical Systems (CUDOS), MQ Photonics Research Centre, Department of Physics and Astronomy, Macquarie University, Sydney, Australia
* yudin@fotonik.dtu.dk

## Abstract

Space division multiplexing using multicore fibers is becoming a more and more promising technology. In space-division multiplexing fiber network, the reconfigurable switch is one of the most critical components in network nodes. In this paper we for the first time demonstrate reconfigurable space-division multiplexing switching using silicon photonic integrated circuit, which is fabricated on a novel silicon-on-insulator platform with buried Al mirror. The silicon photonic integrated circuit is composed of a 7x7 switch and low loss grating coupler array based multicore fiber couplers. Thanks to the Al mirror, grating couplers with ultra-low coupling loss with optical multicore fibers is achieved. The lowest total insertion loss of the silicon integrated circuit is as low as 4.5 dB, with low crosstalk lower than -30 dB. Excellent performances in terms of low insertion loss and low crosstalk are obtained for the whole C-band. 1 Tb/s/core transmission over a 2-km 7-core fiber and space-division multiplexing switching is demonstrated successfully. Bit error rate performance below $10^{-9}$ is obtained for all spatial channels with low power penalty. The proposed design can be easily upgraded to reconfigurable optical add/drop multiplexer capable of switching several multicore fibers.

## Introduction

The communication capacity over standard single mode fibers has been approaching the theoretical limit [1]. In order to further increase the communication capacity over fibers to satisfy the huge capacity demand in future, space division multiplexing (SDM) using multicore fibers (MCFs) has shown to be a promising technology [2-5]. In order to facilitate the deployment of SDM technologies, it is crucial to develop devices that can provide additional network functionalities, such as fan-in/fan-out devices [6-9], MCF amplifiers [10, 11], add/drop modules and switches [12-15], etc. MCF switching is one of the most important functionalities at reconfigurable optical add/drop multiplexers (ROADMs) in future SDM systems. Recently, an array of wavelength-selective switches (WSSs) has also been proposed to implement a ROADM for MCFs communication systems [16]. Flexible architectures using free-space switching based on Micro Electro Mechanical Systems (MEMS) mirrors or on Liquid Crystal on Silicon (LCOS) pixel arrays have also been demonstrated [16, 17]. In addition, all-optical nonlinear switching in MCFs has also been demonstrated using high-power ultrashort laser pulses [18]. Nevertheless, all these solutions are quite complex with high insertion losses. In order to reduce the insertion loss, flexural acoustic waves have also been used for switching in MCFs [19]. However, all current demonstrations are non-integration solutions, and integrating multiple functionalities, as MCF couplers, MCF switches, add/drop on a single platform such as silicon for an MCF system is highly desirable. Integration is a very promising solution because it can take advantage of complementary metal–oxide–semiconductor (CMOS) compatible fabrication process and support massive production, resulting in ultra-compact and powerful silicon chips for MCF systems with potentially low-cost.

In this paper, we for the first time demonstrate reconfigurable switching between the cores of an MCF using a novel silicon photonic integrated circuit (PIC), which is fabricated on a silicon-on-insulator platform with buried Al mirror. The silicon PIC integrates grating coupler array based MCF couplers and a 7x7 switch. Thanks to the Al mirror, ultra-high coupling efficiency between the chip and MCFs is achieved, and the insertion loss of the silicon PIC is as low as 4.5 dB, with low channel dependent loss lower than 2.5 dB and low crosstalk lower than -35 dB.

The good performances in terms of low insertion loss and low crosstalk are obtained for the whole C-band. We further demonstrate core switching of the MCFs in different switching configurations when each core is carrying independent 1 Tb/s data. Bit error rate (BER) performance lower than $10^{-9}$ is achieved for switching all the cores with low power penalty. We further show that the proposed design can be easily upgraded by adding more MZIs into the switching matrix for switching of multicore fibers and realization of more complex MCF ROADMs.

## Results

### 1. Reconfigurable silicon PIC

The topology structure of the silicon PIC is depicted in Fig. 1, which consists of MCF coupler for MCF input and output respectively, and a reconfigurable 7x7 switch. A seven-core fiber is coupled to the input MCF coupler, which is further connected to the 7x7 switch composed of a Mach-Zehnder interferometer (MZI) arrays based switching matrix. After switching, the seven spatial channels are coupled to the output MCF through a second MCF coupler. In this scheme, any core of the input MCF can be reconfigurably switched to any core of the output MCF fiber. For example, by configuring the corresponding MZIs, the bar (solid line) switching configuration for the seven spatial channels (corresponding to different color) can be replaced by the cross (dash line) configuration.

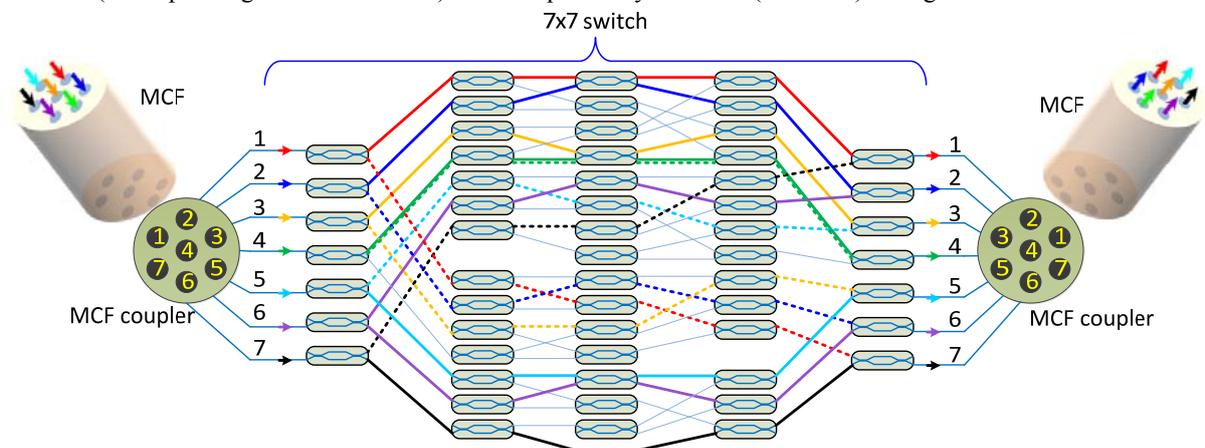

**Fig. 1.** Topology of the silicon PIC for core switch of MCFs. A seven-core fiber is coupled to the input MCF coupler, which is connected to the MZI arrays based 7x7 switching matrix. After switching, the seven spatial channels are coupled to the output MCF through a second MCF coupler. By configuring the corresponding MZIs, the bar (solid line) switching configuration for the seven spatial channels can be freely tuned to cross (dash line) configuration.

The whole silicon PIC is designed on a SOI platform with top silicon thickness of 250 nm. In order to simplify the fabrication process, single etched grating couplers based on photonic crystals [6] are used for the MCF couplers [20]. From the topology of the silicon PIC, building blocks such as 2x2 multimode interferometers (MMIs) and cross intersections are used, and low insertion loss of those building blocks are critical to achieve low insertion loss of the whole silicon PIC. In order to fabricate the whole silicon PIC in the same process, the 2x2 MMIs and cross intersection are designed by three dimensional (3D) finite-difference time-domain method (FDTD) for single etch process. The cross intersection is based on two crossed MMIs, and the crosstalk is minimized by designing the self-image position on the cross point of the two MMIs [21]. In order to achieve ultra-high coupling efficiency, Al mirror is used below the grating coupler based MCF couplers. For this purpose, a new silicon-on-insulator platform with Al mirror dedicated for passive silicon photonics was fabricated first by flip-bonding method. The silicon dioxide layer between the top silicon layer and Al mirror is designed to be 1.6 μm, which is an optimum thickness for fully etched grating couplers on 250 nm silicon layer [20]. The silicon PIC was then fabricated on the new silicon-on-insulator platform by a single step of standard SOI processing, including e-beam lithography and inductively coupled plasma (ICP) etching, followed by extra steps of lithography and metal liftoff for fabrication of metallic heater. Figure 2(a) shows the fabricated silicon PIC, which is wire-bonded to a PCB board for flexible controlling by microprocessor. The detailed microscopy image of the silicon PIC is shown in Fig. 2(b). The 7x7 switch is built out by 57 Mach-Zehnder interferometric (MZI) structures, each incorporating a heater in one arm. Apodized grating coupler array with layout that corresponds to the cores of the MCF is used for MCF coupling [6, 18], as shown in Fig. 2(c). The reconfigurable switching is realized by a MZI matrix, where the tunability is realized by titanium (Ti) heaters [22]. Fig. 2(d) and 2(e) show the fabricated 2x2 MMI and cross intersection used in the silicon PIC, respectively.

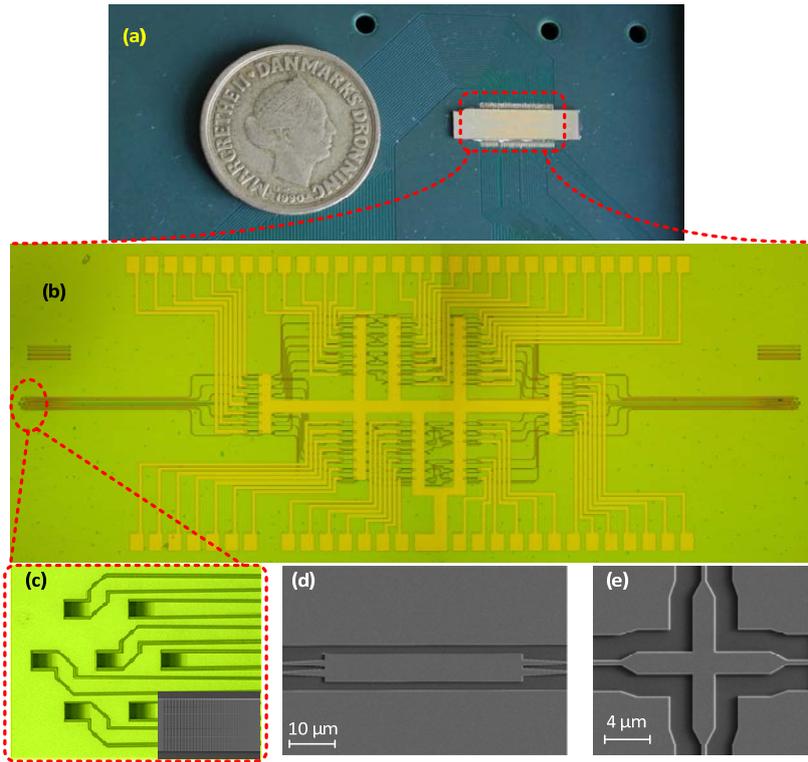

**Fig. 2.** (a) Fabricated silicon PIC, which is wire-bonded to a PCB board. (b) Detailed microscopy image of the silicon PIC, which consists of (c) grating coupler array based MCF coupler, and switching matrix. The reconfigurability is realized by thermal tuning of heaters in the switching matrix. The inset of (c) shows the scanning electron microscopy (SEM) image of the apodized grating coupler. (d) and (e) shows the SEM images of the 2x2 MMI and cross intersection used in the PIC.

The performance of the 2x2 MMI and heaters that are used in the PIC are tested by an asymmetrical MZI (AMZI) with the same heater design, as presented in Fig. 3(a). An extinction ratio larger than 35 dB with insertion loss as low as -0.04 dB is achieved indicating that similar insertion loss and extinction ratio will be obtained for a single MZI switch. In addition, around 5 V (corresponding to 25 mW) results in a phase shift of $2\pi$. A square waveform is applied to test the switching speed of the heater, and typical slow rising time (10% to 90%) and falling time (90% to 10%) of 66 μs and 27 μs are found, respectively. The performance of the cross intersection is tested by cascading different number $N$ of intersections, and measuring the transmitted power. The results are exhibited in Fig. 3(b). An ultra-low insertion loss lower than 0.1 dB is achieved with ultra-low crosstalk lower than -45 dB.

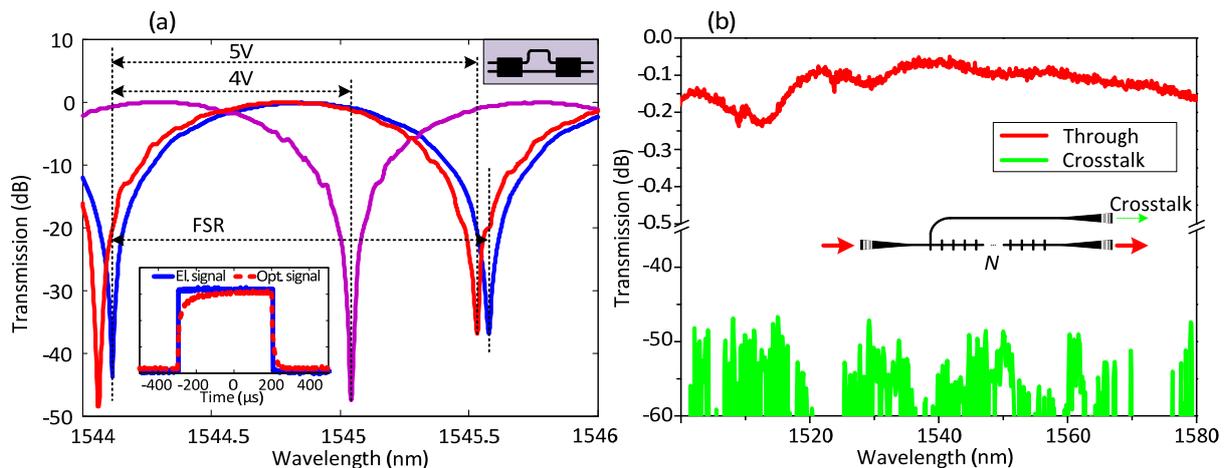

**Fig. 3.** (a) Characterization of AMZI (the upper-right inset) by applying different voltage in order to test performance the MZI switches and heaters. The lower-down inset shows the switched optical signal by electrical square waveform, indicating a switching time of 66 μs and 27 μs for rising and falling time. (b) Characterization of cross intersection by cascading different number of cross intersections, and measuring the corresponding crosstalk arm and transmitted power.

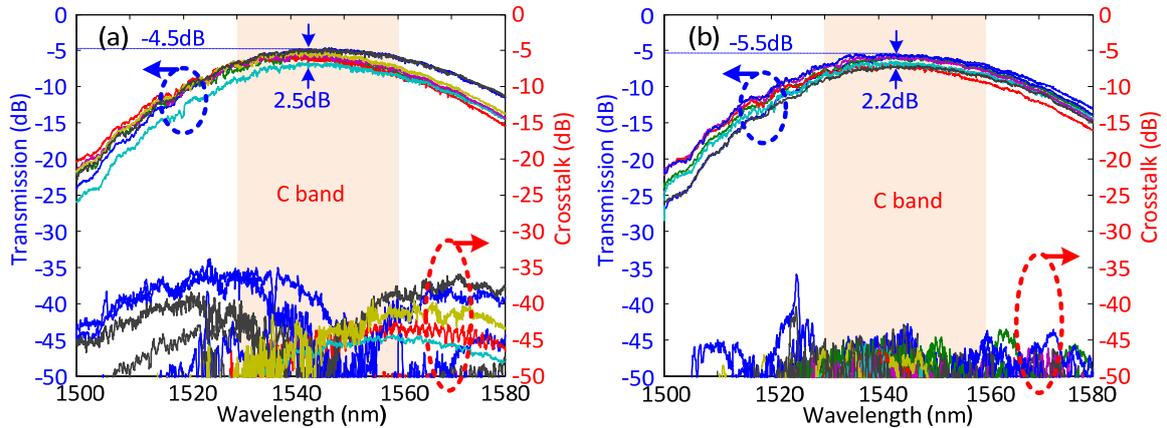

**Fig. 4.** Measured transmission and corresponding crosstalk for (a) bar and (b) cross switching configurations, showing low insertion loss with low channel dependent insertion loss and high extinction ratio to crosstalk over the C band for both configurations.

The whole silicon PIC is then tested by coupling light to different spatial channels in the input MCF through corresponding input grating couplers, and measuring the output from different spatial channels from the corresponding output grating couplers. Fig. 4(a) shows the transmission and corresponding crosstalk for bar configuration (core 1 at the input is connected to core 1 at the output, core 2 to core 2, etc.). High transmission covering the whole C-band is obtained for all the switching paths with crosstalk lower than -30 dB. An extremely low insertion loss of 4.5 dB with channel dependent loss lower than 2.5 dB is achieved. It should be noted that the insertion loss includes the coupling loss of the input and output MCF couplers, the waveguide propagation loss, and all losses by the MZIs and cross intersections. By applying proper voltages to the corresponding heaters in the 7x7 switching matrix, the silicon PIC is tuned to cross switching configuration (core 1 at the input is connected to core 7 at the output, core 2 to core 6, etc.). High transmission is still obtained for all switching paths with lowest insertion loss of 5.5 dB and 2.2 dB channel dependent loss. Low crosstalk (lower than -35 dB) is obtained for the whole C-band.

### 2. System experiment

Fig. 5 shows the system experimental setup. A wavelength division multiplexing (WDM) signal in the range from 1541.35 nm to 1560.61 nm consisting of 25 channels with a 100 GHz grid is used. Each channel is carrying 40 Gb/s on-off keying (OOK) modulated data, resulting in 1 Tb/s/channel traffic load. After modulation, the data is split to seven spatial channels, amplified and decorrelated using fibers with different length for each channel before loading the seven-core fiber. An on-chip fan-in device based on 3D inscribed waveguides is used for coupling to a 2-km seven-core fiber. After propagation, the seven-core fiber is coupled directly to the fabricated silicon PIC for spatial channel switching. The 7x7 switch, which is able to switch the different cores of the MCF, is controlled by a microcontroller allowing for individual control of each heater on the chip. A preliminary characterization is performed to determine the optimum values of the relevant heaters in order to establish data paths in bar and cross configurations. In order to receive the data from the switched cores at the output of the switch, an additional integrated fan-out device is used for spatial demultiplexing after which the data is amplified, wavelength de-multiplexed using a flat-top bandpass 100 GHz filter and passed to a 40 Gb/s receiver.

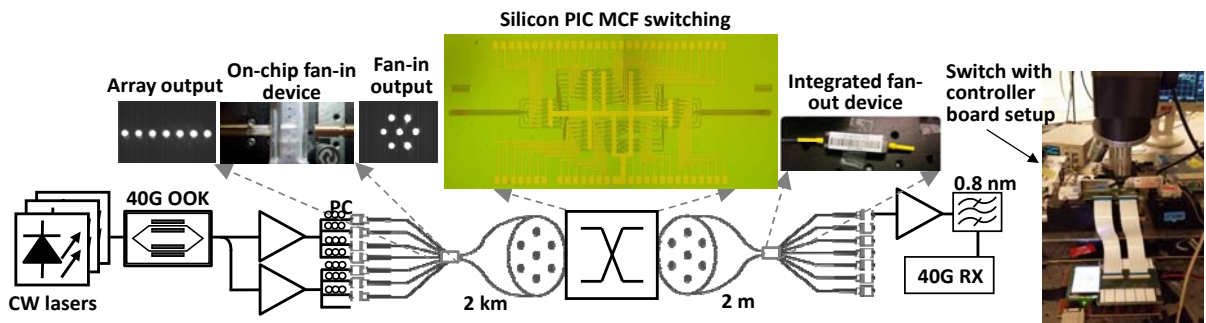

**Fig. 5.** Experimental setup. 1 Tb/s OOK traffic load covering the C band is used as data. The data is split to seven beams (channels), amplified and decorrelated using fibers with different lengths. The seven channels are coupled to a 2-km seven-core fiber by a 3D inscribed waveguides based on-chip fan-in device. After transmission, the seven channels are coupled to the chip for switching. The switched seven spatial channels are coupled out to a 2 m seven-core fiber again, and an additional integrated fan-out device is used for spatial demultiplexing after which the data is amplified, wavelength de-multiplexed and received in a 40 Gb/s receiver

Figs. 6a and 6b illustrate the output spectra of all the cores after the switch in the bar and cross configurations respectively. The performance of all channels in a single core in both configurations is shown in Fig. 6c. In both cases all channels exhibit similar performance with an average penalty of 3 to 4 dB compared to the average back-to-back (B2B) performance. The imperfect power equalization of the channels as well as the wavelength dependent crosstalk of the switch, which is due to polarization variation after transmission, contributes to the penalty variations. In order to confirm that similar performance is expected in all the switched cores, the receiver sensitivity (BER=$10^{-9}$) of one channel (1550.92 nm) is measured in all cores as shown in Fig. 6d. For both configurations, the performance in all cores is within 5 dB margin. The different insertion loss and crosstalk experienced by each core as a result of the coupling devices as well as the different crosstalk in the switching configurations contribute to this variation. The effect of crosstalk from switching can be seen as 5 dB variation in the sensitivity of the chosen channel in core 4, which in both configurations is switched in the same way. In order to precisely evaluate the impact of the crosstalk from switching, full BER curves are measured on the same channel in both configurations as shown in Fig. 6e and Fig. 6f. First, data is launched in only one core of the MCF and switched (i.e. w/o crosstalk) and then data is launched and switched in all cores simultaneously (i.e. w/ crosstalk). There is around 1.5 dB and 1.2 dB penalty when only one core is loaded for bar and cross configurations respectively, which is due to crosstalk from other coupling devices used in the system. Loading and switching all cores results in about 0.4 dB and 2.9 dB additional penalty for bar and cross configurations, respectively. This indicates that some cores may suffer from higher crosstalk in some configurations.

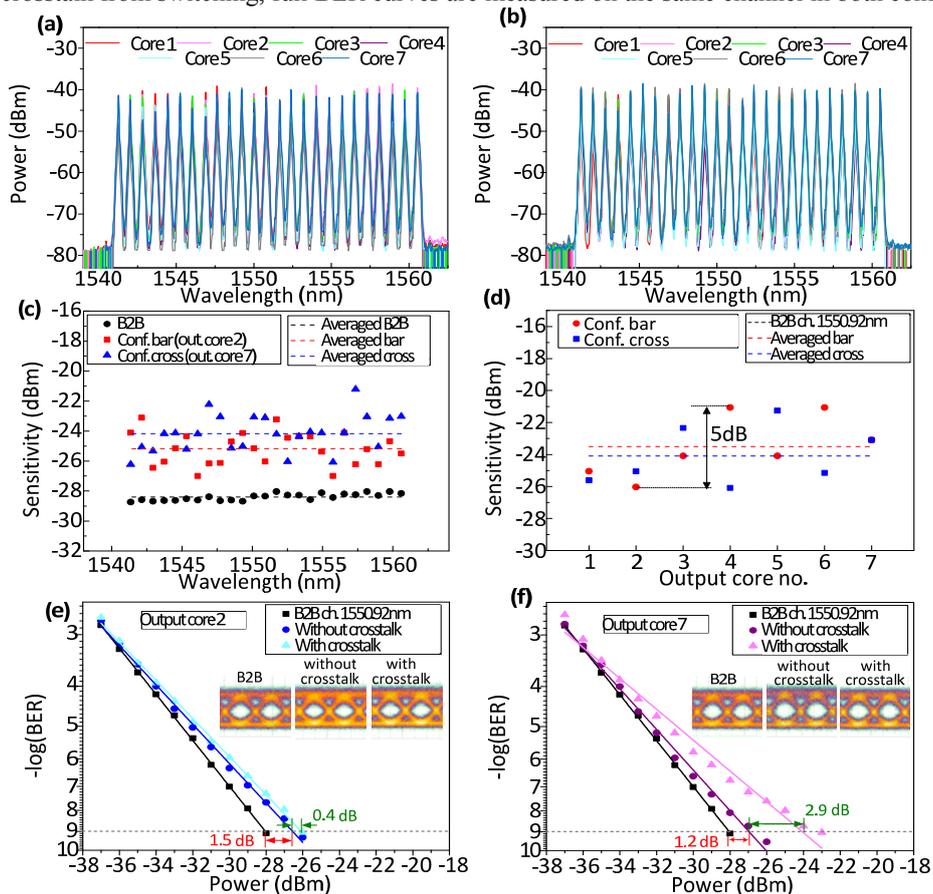

**Figure 6.** Output spectra of the switched cores in (a) configuration bar and (b) configuration cross; (c) receiver sensitivity of all channels in a single core in both configurations and (d) receiver sensitivity of a single channel (1550.92 nm) in all cores in both configurations; BER results and eye diagrams of a single channel (1550.92 nm) in a single core w/o and w/ crosstalk in (e) configuration bar and (f) configuration cross.

## Discussion and Conclusions

It is observed that a single AMZI has an extinction ratio larger than 35 dB, indicating that similar extinction ratio should be obtained for MZIs in the switching matrix. The switching matrix of the silicon PIC is realized by cascading several MZIs, and as a result, an extinction ratio much larger than 35 dB to the spatial crosstalk should be obtained after the switch. The measured crosstalk of the silicon PIC is -30 dB, indicating that the heating power for some MZIs in the switching matrix may not be well optimized. Thus, further careful optimization of the heating power will greatly reduce the crosstalk, and therefore reduce the power penalty accordingly. In addition, polarization variation results in wavelength dependent sensitivity that can be greatly improved by polarization diversity technology [23-26]. The power consumption and switching speed are very important for optical fiber network nodes.

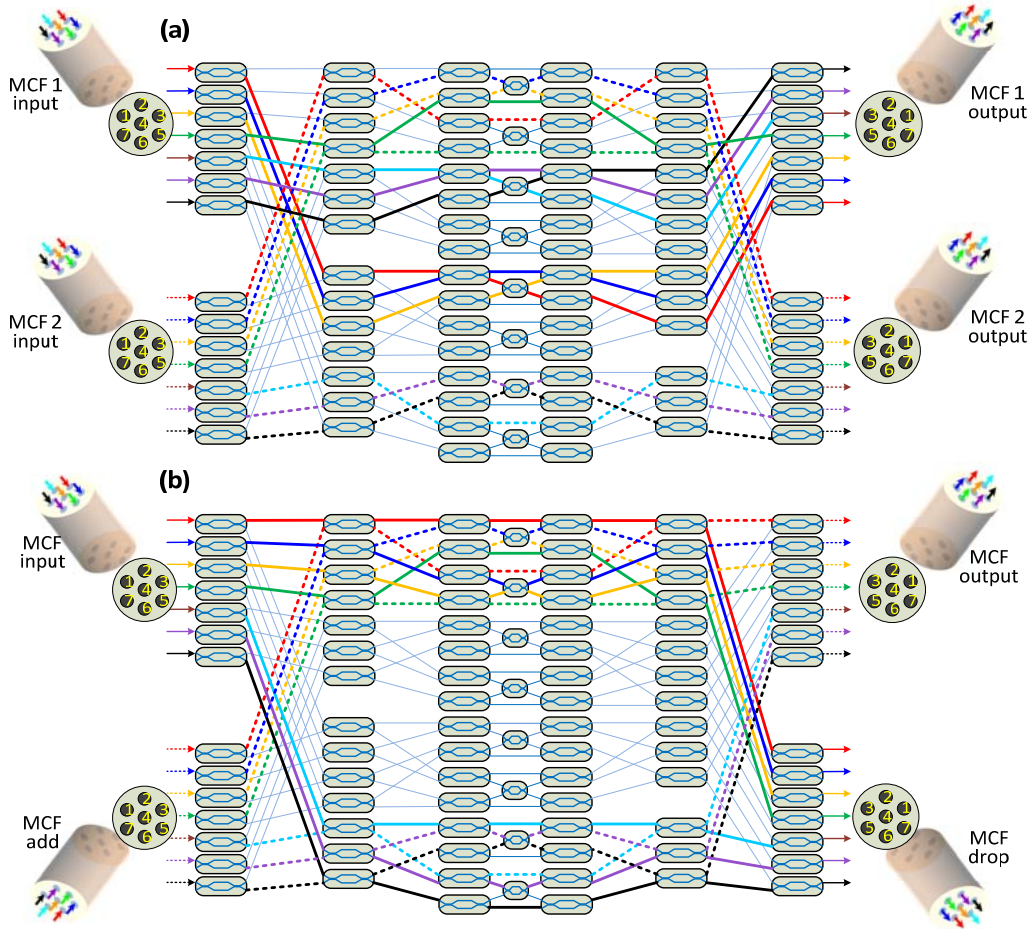

**Fig. 7.** (a) Topology of a silicon switch device for switching of two MCFs. MCF 1 is switched to the output MCF 1 by cross configuration through solid line routing paths, and at the same time, MCF 2 is switched to output MCF 2 by bar configuration through dash line routing paths. (b) The device can also be used as a ROADM for MCFs. The seven channels of the input MCF can be dropped to the MCF drop output through solid routing paths, and the data can be added simultaneously from an add MCF to the MCF output through dash routing paths.

In this work, the switching power for a single MZI is not very low, due to that a thick silicon dioxide layer is introduced between the silicon waveguide and the Ti heaters to avoid metallic loss. More efficient heater designs [27-30] can be used to reduce the power consumption. The low switching speed in the current work may be attributed to the thick BCB bonding layer, which has relatively low thermal conductivity. The switching time can be improved by thinning down the BCB layer. On the other hand, improved design that directly integrates thermal heaters on silicon waveguide is a promising solution to provide fast switching speed [31]. Other heater materials, such as graphene can also be used as an efficient heater with fast switching time [32].

The current device can be upgraded to switch two MCFs by adding more MZIs into the switching matrix, as shown in Fig. 7(a). MCF 1 and MCF 2 can be freely switched and routed to the corresponding output MCF. As an example, MCF 1 is switched to MCF 1 output on cross configuration through the solid line routing paths, and at the same time, MCF 2 is switched to MCF 2 output on bar configuration through the dash line routing paths. The same upgraded device can also be used as a real MCF ROADM module. As shown in Fig. 7(b), seven spatial channels from the input MCF can be dropped to the MCF drop output through solid line routing paths, and data can be added from the adding MCF to the MCF output through the dash line routing paths.

In summary, we have designed and fabricated a novel silicon PIC for MCF switching. The silicon PIC is fabricated on a novel SOI platform with Al mirror. Low insertion loss with high extinction ratio and low crosstalk is obtained over a large bandwidth. We have demonstrated BER performance lower than $10^{-9}$ when transmitting 1 Tb/s/core in a 2-km MCF and switching all the cores in different configurations using the silicon PIC, constituting a first demonstration of an integrated device with this functionality. The proposed device can be upgraded to support additional features such as switching several MCFs and thus providing true MCF ROADM functionalities, which will be extremely useful for future MCF networks.

## Methods

**Silicon-on-insulator wafer with Al mirror**

The silicon-on-insulator wafer with buried Al mirror was fabricated by flip-bonding method. A silicon dioxide layer with optimum thickness of 1.6 μm was first deposited by plasma-enhanced chemical vapor deposition (PECVD) on the commercial SOI wafer, which has 250 nm thick top silicon layer on 3 μm thick buried oxide layer. Al mirror was sputtered by e-beam afterwards. After that, a second layer of silicon dioxide with 500 nm thickness was deposited. The SOI wafer and a carrier silicon wafer are then both spun with 500 nm benzocyclobutene (BCB) layer, and the SOI wafer was then flip-bonded to the carrier silicon wafer. The bonded wafers are put into an oven for BCB hard curing. The substrate of the SOI wafer was then removed by dry etching. The new silicon-on-insulator with Al mirror was finally obtained by buffered hydrofluoric acid (BHF) wet etching of the buried oxide layer of the original SOI wafer.

**Silicon PIC fabrication**

The silicon PIC circuit was fabricated on the new silicon-on-insulator wafer with Al mirror. A single step of standard SOI processing, including e-beam lithography and inductively coupled plasma (ICP) etching was first used to fabricate the whole silicon PIC simultaneously. A 1500 nm thick layer of $SiO_2$ was then deposited on top of the chip. The chip surface was then polished, and the top $SiO_2$ was thinned down to 1 μm accordingly. The 1 μm is used as an isolation layer from the Ti heaters fabricated later to avoid potential optical losses. Afterwards, the 100 nm thick titanium heaters are formed by e-beam lithography followed by metal deposition and liftoff process. Then the thick Au/Ti contact layer was fabricated by UV lithography followed by metal deposition and liftoff process. The chip was then cleaved and wire-bonded to a PCB board for test.

## Acknowledgements

This work is supported by the Danish Council for Independent Research (DFF-1337-00152 and DFF-1335-00771) and the ECFP7 grant no. 619572, COSIGN. We would like to thank Optoscribe for providing the integrated fan-out device.